\documentclass[aps,twocolumn,prb,superscriptaddress,longbibliography]{revtex4-2}
\usepackage[colorlinks,breaklinks,citecolor=blue,linkcolor=blue,urlcolor={blue}]{hyperref}
\usepackage[intlimits,tbtags]{amsmath}
\usepackage{times}
\usepackage{color}
\usepackage{amssymb}
\usepackage{graphicx}
\usepackage{bm}
\usepackage{array}
\usepackage{dcolumn}
\usepackage{bm}
\usepackage{setspace}
\usepackage[T1]{fontenc}

\usepackage{xcolor}

\newcommand{\wuli}{School of Physics, Harbin Institute of Technology, Harbin, 150001, China}
\newcommand{\dengliziti}{Heilongjiang Provincial Key Laboratory of Plasma Physics and Application Technology, Harbin Institute of Technology, Harbin 150001, China}
\newcommand{\nengyuan}{School of Energy Science and Engineering, Harbin Institute of Technology, Harbin 150001, China}

\begin{document}
\title{Guest metal-driven quantum anharmonic effects on stability and two-gap superconductivity in carbon-boron clathrates}

\author{Xianghui Meng}
\affiliation{\wuli}

\author{Yanqing Shen}\email{shenyanqing2004@163.com}
\affiliation{\wuli}

\author{Xin Yang}
\affiliation{\wuli}

\author{Xinyu Wang}
\affiliation{\wuli}

\author{Qing Ai}\email{hitaiqing@hit.edu.cn}
\affiliation{\nengyuan}

\author{Yong Shuai}\email{shuaiyong@hit.edu.cn}
\affiliation{\nengyuan}

\author{Zhongxiang Zhou}\email{zhouzx@hit.edu.cn}
\affiliation{\wuli}
\affiliation{\dengliziti}

\date{\today}
\begin{abstract}
Traditionally, strong quantum anharmonic effects have been considered a characteristic of hydrogen-rich compounds.
Here we propose that these effects also play a decisive role in boron–carbon clathrates. 
The stability and superconducting transition temperature ($T_\text{c}$) of carbon–boron clathrates XYB\textsubscript{6}C\textsubscript{6}, whose metal atoms have an average oxidation state of +1.5, have long remained under debate. 
At this oxidation state, some combinations (e.g., RbSrB\textsubscript{6}C\textsubscript{6}) are dynamically stable, whereas others (e.g., RbPbB\textsubscript{6}C\textsubscript{6}) are not. 
Using the stochastic self-consistent harmonic approximation combined with machine learning, we find that the anharmonicity originates primarily from guest metal atoms. 
For comparison, we find that quantum fluctuations have negligible influence on SrB\textsubscript{3}C\textsubscript{3} but remove the lattice instability of RbPbB\textsubscript{6}C\textsubscript{6}. 
The predicted $T_\text{c}$ of RbPbB\textsubscript{6}C\textsubscript{6} ($\approx$ 88~K) is nearly twice that of SrB\textsubscript{3}C\textsubscript{3}. 
Moreover, RbPbB$_6$C$_6$ exhibits two-gap superconductivity due to the higher C/B ratio in the density of states at the Fermi level compared to SrB$_3$C$_3$, weakening the $sp$\textsuperscript{3} hybridization.
These findings demonstrate that quantum anharmonicity crucially governs the stability and superconductivity of XYB\textsubscript{6}C\textsubscript{6} clathrates.\end{abstract}
\maketitle

\section{Introduction}
Light-element-based materials have recently emerged as promising candidates for phonon-mediated superconductivity under ambient conditions, owing to their strong electron-phonon coupling (EPC) and tunable electronic structures~\cite{ref1,ref2,ref3,ref4,ref5}. 
The identification of many boron-carbon (B-C) materials has been primarily driven by theoretical predictions~\cite{ref6,ref7}, similar to the discovery process of high-pressure hydrides (e.g., H\textsubscript{3}S, CaH\textsubscript{6}, LaH\textsubscript{10}, LaSc\textsubscript{2}H\textsubscript{24}), which preceded experimental methods~\cite{ref8,ref9,ref10,ref11}. 
This research paradigm emphasizes the potential of first-principles methods, including crystal structure prediction (CSP) searches and EPC calculations~\cite{ref12,ref13,ref14}, as invaluable tools to identify promising superconductors for future synthesis. 
A class of carbon-boron framework, initially pinpointed theoretically using CSP searches, has been synthesized in a host/guest clathrate structure~\cite{ref6,ref7}. 
The SrB\textsubscript{3}C\textsubscript{3} clathrate adopts the bipartite sodalite structure, where host cages are $sp^3$-bonded truncated octahedral C\textsubscript{12}B\textsubscript{12} framework that encapsulate Sr guest atoms. 
Among the first 57 elements of the periodic table (X = H–La), only Ca, Sr, Y, Ba, and La are capable of forming stable XB\textsubscript{3}C\textsubscript{3} compounds under ambient pressure, with the critical temperature $T_\text{c}$ consistently remaining below 50 K~\cite{ref15}.

A key goal is to align the intrinsic properties of such materials with BCS theory (e.g., high density of states near the Fermi level, strong electron-phonon coupling, and high-frequency phonons) to address their low $T_\text{c}$ issue~\cite{ref16}. 
Recent theoretical predictions for XYB\textsubscript{6}C\textsubscript{6}, obtained when the metal atoms in XB\textsubscript{3}C\textsubscript{3} are of two different elements, suggest that it could achieve a higher $T_\text{c}$ when the average oxidation state of XY is +1.5~\cite{ref1}. 
However, the dynamical stability of XYB\textsubscript{6}C\textsubscript{6} remains under debate. 
For example, when the average oxidation state of XY is +1.5, XYB\textsubscript{6}C\textsubscript{6} (XY = KSr, KCa, KPb, RbSr) is dynamically stable. 
In contrast, XYB\textsubscript{6}C\textsubscript{6} (XY = RbPb, CsPb) exhibits dynamical instability, especially when one of the X/Y elements is Pb. 
These uncertainties highlight the limitations of classical first-principles methods, particularly when quantum fluctuations significantly influence material behavior.

Quantum fluctuations raise concerns about the applicability of classical first-principles methods in predicting crystal structures and assessing the stability of hydrides~\cite{ref17,ref18,ref19}. 
Experimental results confirm that the Fm\textsubscript{3}m structure of LaH\textsubscript{10} is the ground state at pressures between 137 and 218 GPa, contradicting theoretical predictions of lattice distortion~\cite{ref20}. 
Like LaH\textsubscript{10}, XYB\textsubscript{6}C\textsubscript{6}, with a sodalite-type structure with Pb as the X/Y element, may encounter similar stability issues. 
However, the stochastic self-consistent harmonic approximation (SSCHA) method used to simulate quantum fluctuations is computationally demanding~\cite{ref21}. 
Therefore, it is essential to introduce a new approach that combines SSCHA with machine learning (ML) techniques, establishing a novel standard for evaluating the stability of such materials.

In this work, we combine SSCHA with ML methods to investigate quantum anharmonicity in XYB\textsubscript{6}C\textsubscript{6}. 
Specifically, quantum anharmonicity is closely related to the metal guest atoms in XYB\textsubscript{6}C\textsubscript{6} and is particularly important for XYB\textsubscript{6}C\textsubscript{6} with high EPC constants. 
The effect of anharmonicity on the dynamical stability and $T_\text{c}$ of SrB\textsubscript{3}C\textsubscript{3} is negligible. 
Anharmonicity is primarily driven by the heavy atom Pb, and we illustrate this using RbPbB\textsubscript{6}C\textsubscript{6} as an example. 
After considering quantum anharmonic effects, the dynamical stability of RbPbB\textsubscript{6}C\textsubscript{6} is significantly improved. 
By solving the anisotropic Migdal-Eliashberg equations, we further predict an EPC constant of 2.87 for RbPbB\textsubscript{6}C\textsubscript{6} and calculate its $T_\text{c}$ to be approximately 88 K, indicating its potential for high-temperature superconductivity.

\section{COMPUTATIONAL DETAILS}
First-principles calculations were conducted using the QUANTUM ESPRESSO package~\cite{ref22}, implementing the generalized gradient approximation (GGA) as formulated by Perdew-Burke-Ernzerhof~\cite{ref23}. 
The energy cutoffs were set at 80 Ry for wavefunctions and 320 Ry for charge density and potential, employing optimized Norm-Conserving Vanderbilt pseudopotentials~\cite{ref24,ref25}. 
Structural optimizations and self-consistent calculations were performed on a 10 × 10 × 10 k-point grid within the Brillouin zone, adhering to stringent convergence criteria: 10\textsuperscript{-6} Ry for total energy and forces during ionic relaxation and 10\textsuperscript{-10} Ry for electronic convergence. 
Harmonic phonon properties were calculated using density functional perturbation theory on a 5 × 5 × 5 q-point mesh..

The quantum zero-point motion and anharmonic effects were evaluated within the SSCHA framework~\cite{ref26}.
To minimize the free energy, a finer stochastic sampling of 1000 configurations was employed at each iteration~\cite{ref27,ref28}. 
A 2 × 2 × 2 supercell containing 112 atoms was utilized to calculate the energy, forces, and stress, with the Monkhorst–Pack k-point grid scaled to 6 × 6 × 6. 
To accelerate the prediction of energy, forces, and stress for each configuration, first-principles methods were combined with machine learning techniques. 
Machine learning interatomic potentials (MLIPs) were trained using a randomly selected training set of 200 structures and used to predict energy, forces, and stress for finer stochastic sampling in subsequent iterations. 
Anharmonic dynamical matrices were obtained on a commensurate 2 × 2 × 2 q-point grid and interpolated to a 5 × 5 × 5 q-point mesh, based on the difference between the harmonic and anharmonic dynamical matrices in the 2 × 2 × 2 q-point grid.

The anisotropic Migdal-Eliashberg equation was solved using the EPW package, which integrates Maximally Localized Wannier Functions~\cite{ref14,ref29}.
A total of 24 $sp\textsuperscript{3}$-hybridized states were constructed as MLWFs, each with a spatial spread below 1.55 Å. 
The Wannier interpolated results are shown in Figs.~S1 and S2 of the Supplemental Material (SM)~\cite{refSM}. 
Electron-phonon matrix elements were interpolated over a dense 60 × 60 × 60 k-point grid and a 20 × 20 × 20 q-point grid. 
A Fermi surface window of 0.4 eV was selected to evaluate the states in the self-energy delta functions. 
Visualization of the Fermi surface was carried out using the FERMISURFER package~\cite{ref30}.
	
\section{RESULTS AND DISCUSSION}
\begin{figure}[htp]
	\centering
	\includegraphics[width=1\linewidth,angle=0]{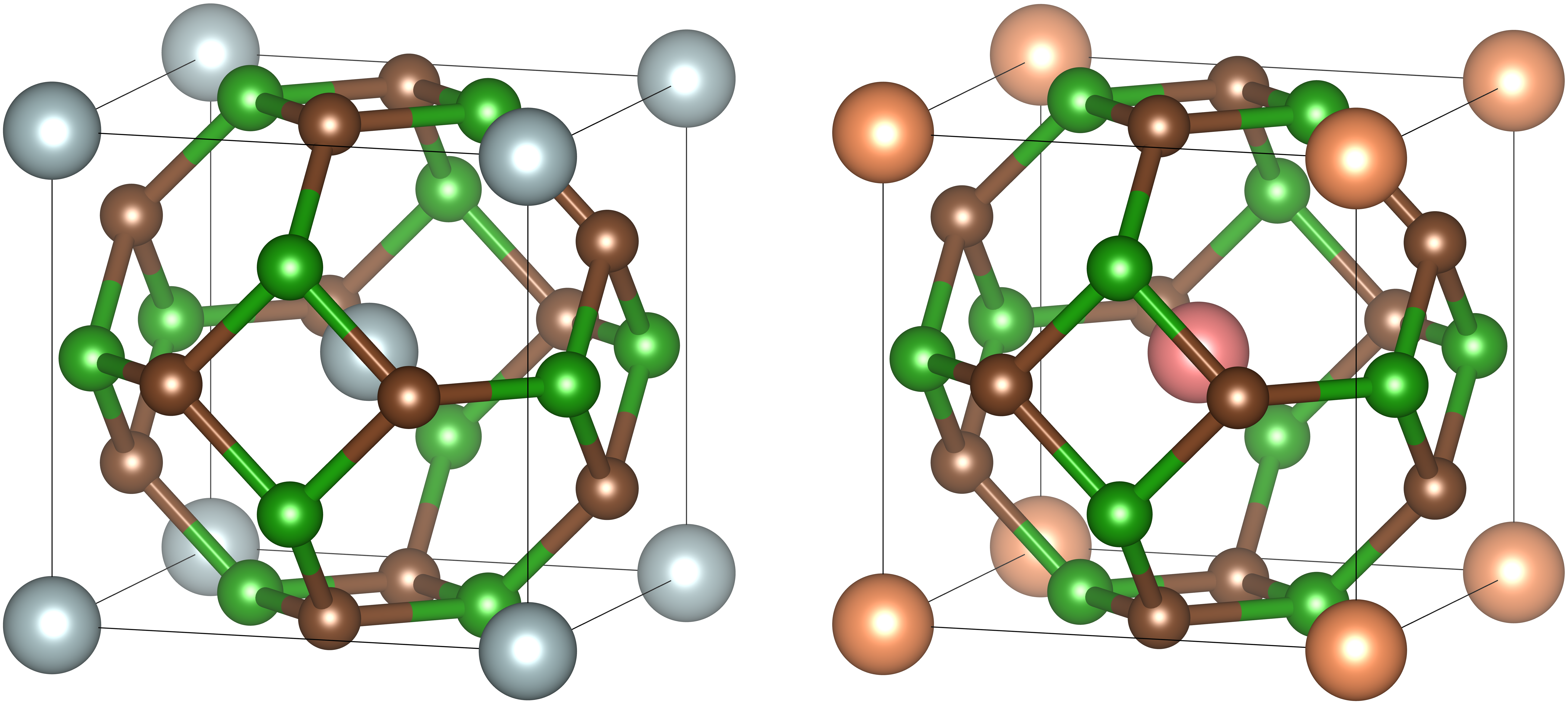}
	\caption{Crystal structures for (a) SrB$_3$C$_3$ and (b) RbPbB$_6$C$_6$. The atoms are represented by silver, orange, red, green, and brown spheres, corresponding to Sr, Rb, Pb, B, and C, respectively.}
	\label{fig:cs}
\end{figure}

Binary-guest C\textendash B clathrates XYB\textsubscript{6}C\textsubscript{6} are found to exist using CSP searches, allowing X $\neq$ Y~\cite{ref1,ref6}.
Similar to LaH\textsubscript{10}~\cite{ref20,ref31}, quantum anharmonic effects play a crucial role in the stability and superconductivity of XYB\textsubscript{6}C\textsubscript{6}, with Pb as the X/Y element. 
As an illustrative example, we consider $sp^3$-bonded carbon-boron clathrates SrB\textsubscript{3}C\textsubscript{3} and RbPbB\textsubscript{6}C\textsubscript{6}. 
Two cubic structures, SrB\textsubscript{3}C\textsubscript{3} and RbPbB\textsubscript{6}C\textsubscript{6}, are depicted in Figure~\ref{fig:cs}, with structural parameters listed in Table S1 of SM~\cite{refSM}. 

As a first step, we assess their mechanical, dynamic, and thermodynamic stability. 
The elastic constants, summarized in Table S2 of the SM~\cite{refSM}, satisfy the Born stability criteria for cubic systems: $C_{11} - C_{12} > 0$, $C_{11} + 2C_{12} > 0$, and $C_{44} > 0$~\cite{ref32}. 
Minimizing the variational free energy requires a substantial number of single-point density functional theory calculations. 
Here, we employ MLIPs to compute forces, stresses, and energies. 
\begin{figure}[htp]
	\centering
	\includegraphics[width=1\linewidth,angle=0]{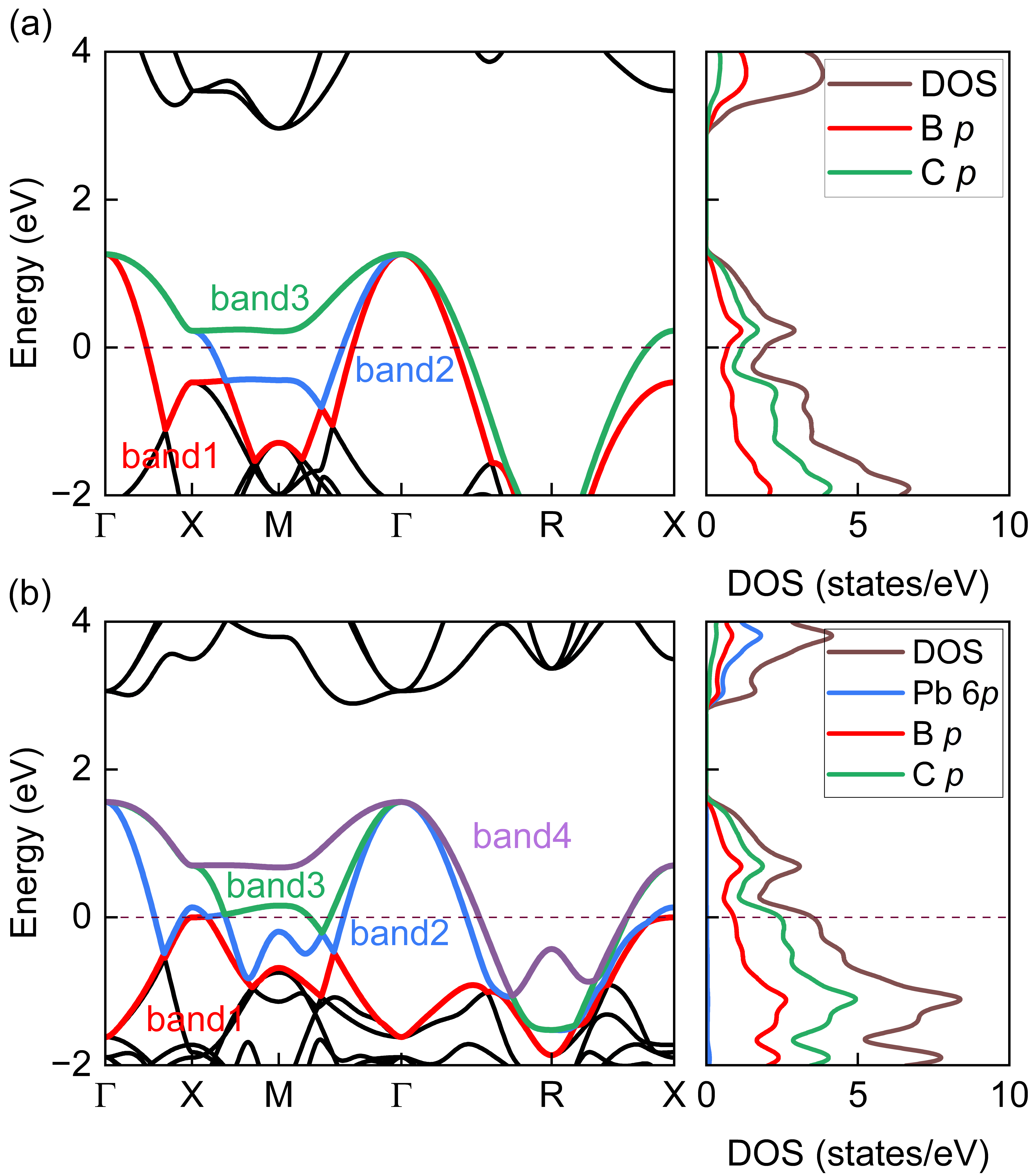}
	\caption{The band structure and projected density of states for (a) SrB$_3$C$_3$ and (b) RbPbB$_6$C$_6$. In band structures, multiple bands dominated by nonmetal atoms that intersect with the Fermi level ($E_\text{F}$ = 0 eV) are marked with different colors.}
    \label{fig:bs}
\end{figure}
\begin{figure*}[htp]
	\centering
	\includegraphics[width=1\linewidth,angle=0]{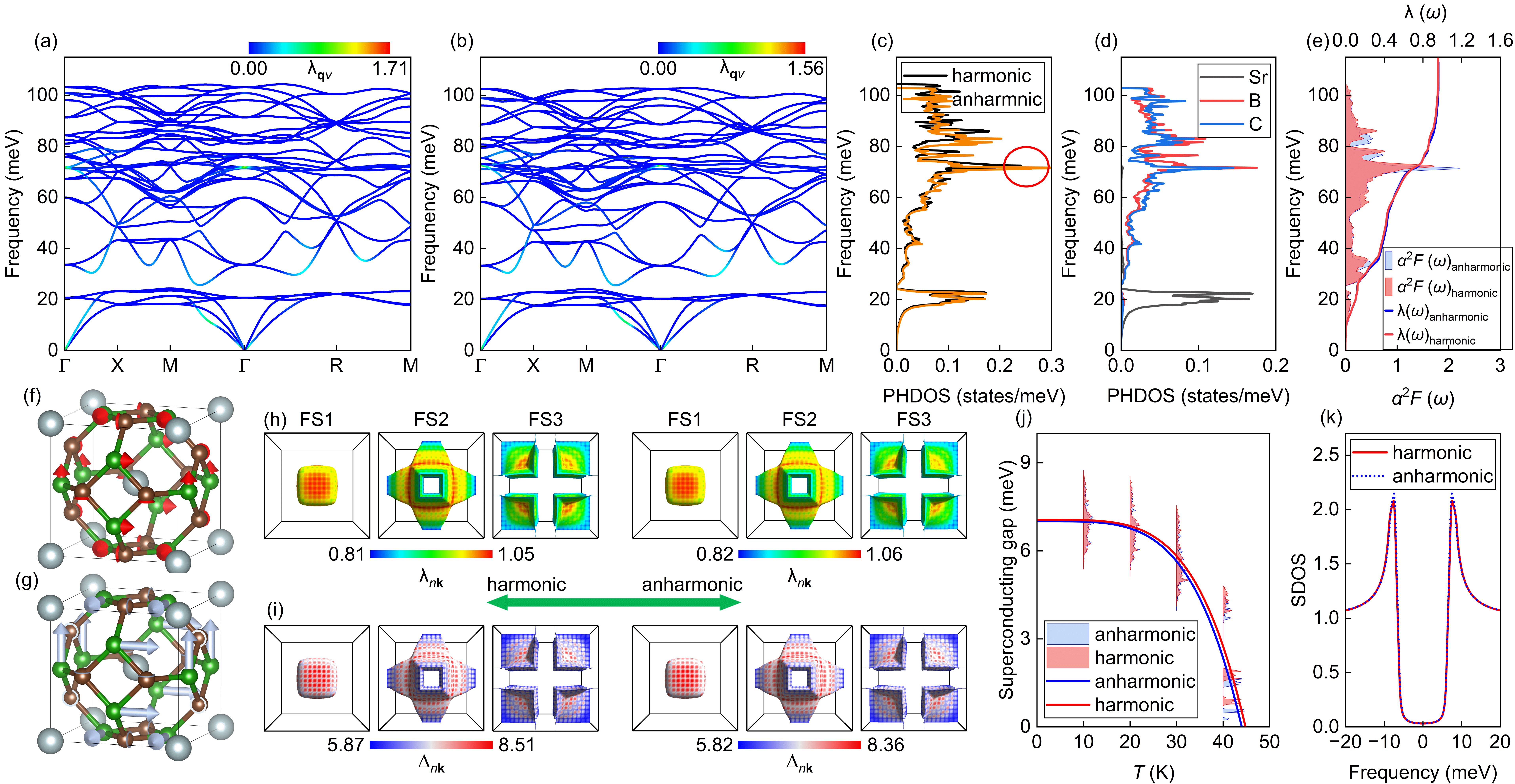}
	\caption{Harmonic and anharmonic phonon properties and EPC of SrB$_3$C$_3$.
	(a) Harmonic phonon spectrum with $\lambda_{q\nu}$. 
    (b) Anharmonic phonon spectrum with $\lambda_{q\nu}$. The color mapping from blue to red in panels (a) and (b) represents the magnitude of $\lambda_{q\nu}$ for each phonon mode, with red indicating the highest values.
    (c) Harmonic and anharmonic phonon density of states (PHDOS). 
    (d) Projected PHDOS on Sr, B, and C atoms in the anharmonic framework. 
    (e) Eliashberg spectral function $\alpha^2F(\omega)$ and cumulative EPC constant $\lambda(\omega)$.
	Double-degenerate $E_g$ modes at the $\Gamma$ point in (f) harmonic and (g) anharmonic frameworks. 
	Displacements are shown by red and light blue arrows for the harmonic and anharmonic frameworks, respectively.
    (h) Momentum-resolved EPC strength $\lambda_{nk}$ on the Fermi surface.
    (i) Superconducting gap $\Delta_{nk}$ on the Fermi surface at 10 K.
    (j) Temperature dependence of the superconducting gap $\Delta$.
    (k) Superconducting density of states (SDOS) at 10 K.
	}
    \label{fig:ph}
\end{figure*}
Figs.~S3 and S4 display the errors in force, energy, and stress calculations produced by MLIPs, which have less influence on SSCHA results due to averaging effects~\cite{refSM}. 
Quantum effects arise from the guest atoms, with their strength dependent on the atomic species. 
After the inclusion of quantum anharmonic effects, the dynamical stability of SrB\textsubscript{3}C\textsubscript{3} remains unchanged, while RbPbB\textsubscript{6}C\textsubscript{6} becomes dynamically stable (see Fig.~S5)~\cite{refSM}. 
SrB\textsubscript{3}C\textsubscript{3} was synthesized at nearly 50 GPa and exists under ambient pressure in inert atmospheres~\cite{ref6}. 
Fig.~S6 shows the computed formation enthalpy (\(\Delta H\)) as functions of pressure for RbPbB\textsubscript{6}C\textsubscript{6}~\cite{refSM}. 
It decomposes into RbC + Pb + 6B + 5C at pressures below 14.5 GPa, which is lower than the theoretical decomposition pressures of other analogous compounds~\cite{ref33}. 
Similarly, RbPbB\textsubscript{6}C\textsubscript{6} may be obtained in a comparable manner as SrB\textsubscript{3}C\textsubscript{3}.

Figure~\ref{fig:bs} presents the electronic band structures and projected density of states (DOS) for SrB\textsubscript{3}C\textsubscript{3} and RbPbB\textsubscript{6}C\textsubscript{6}. 
Three bands in SrB\textsubscript{3}C\textsubscript{3}, primarily derived from nonmetal atoms, intersect with the Fermi level ($E_\text{F}$ = 0 eV) and are labeled with distinct colors (see Fig.~\hyperref[fig:bs]{\ref*{fig:bs}(a)}), in agreement with previous calculations~\cite{ref2,ref15}.
The DOS at the Fermi level plays a decisive role in determining $T_\text{c}$ for this class of materials. 
Tuning the metal-guest atoms enables the alignment of $E_\text{F}$ with regions of higher DOS. 
In RbPbB\textsubscript{6}C\textsubscript{6}, where the average metal oxidation state is +1.5, $E_\text{F}$ shifts downward, crossing additional energy bands and increasing the DOS to 3.5 states/eV (see Fig.~\hyperref[fig:bs]{\ref*{fig:bs}(b)}). 
The states near $E_\text{F}$ and the Fermi-surface pockets are dominated by B~$p$ and C~$p$ orbitals, with the latter contributing more significantly owing to the higher electronegativity of carbon (see Figs.~\ref{fig:bs}, S7, and S8~\cite{refSM}). 
This $p$-orbital–dominated electronic character resembles that of layered boron–carbon intercalated compounds XB\textsubscript{2}C\textsubscript{2} (X = Li, Na, K, etc.)~\cite{ref34,ref35,ref36}. 
The Fermi surface of RbPbB\textsubscript{6}C\textsubscript{6} contains an additional sheet relative to SrB\textsubscript{3}C\textsubscript{3}, originating from the newly occupied bands induced by the downward shift of the Fermi level. 
This additional sheet reflects the enhanced electronic filling and is consistent with the increased DOS at the Fermi level. 

Figures~\hyperref[fig:ph]{\ref*{fig:ph}(a)}-\hyperref[fig:ph]{\ref*{fig:ph}(d)} indicate that quantum anharmonic effects negligibly influence the phonon dispersions, the EPC strength $\lambda_{q\nu}$ and the phonon density of states (PHDOS) for SrB\textsubscript{3}C\textsubscript{3}. 
The anharmonic PHDOS reveals that mid-to-high frequency vibrations are primarily associated with the lighter B and C atoms, while low-frequency modes are dominated by the heavier Sr atoms (see Fig.~\hyperref[fig:ph]{\ref*{fig:ph}(d)}). 
According to anharmonic PHDOS, we can infer that the EPC is dominated by B\textendash C vibrations, similar to cage-like superconductors~\cite{ref37,ref38,ref39}. 
The total EPC constant \(\lambda = 0.96\) remains unchanged between harmonic and anharmonic calculations, in good agreement with previous reports (0.92\textendash 1.02)~\cite{ref2,ref15,ref40,ref41}. 
The cumulative \(\lambda(\omega)\) increases in three distinct steps [Fig.~\hyperref[fig:ph]{\ref*{fig:ph}(e)}]: the first, contributed by acoustic and optical branches 1\textendash 3, accounts for \(\approx 10.8\%\) of \(\lambda\) and reflects strong coupling along the M\textendash \(\Gamma\) path, where the soft acoustic branch couples with electronic states near $E_\text{F}$, similar to Na-intercalated boron carbide~\cite{ref35}; the second step arises from optical branches 7\textendash 9, contributing \(\approx 0.41\) to \(\lambda\); and the third is characterized by a distinct peak in \(\alpha^2 F(\omega)\) at \(\approx 71\) meV, originating from the doubly degenerate $\text{E}_\text{g}$ modes at \(\Gamma\) (see Figs.~\hyperref[fig:ph]{\ref*{fig:ph}(f)} and ~\hyperref[fig:ph]{\ref*{fig:ph}(g)}). 
These in-plane B vibrations within the B\textsubscript{2}C\textsubscript{2} layer yield the largest \(\lambda_{q\nu}\), analogous to the $\text{E}_\text{g}$ modes in CaH\textsubscript{6}~\cite{ref9}.
For comparison, the contributions of each phonon mode to EPC \(\lambda_v\) in both the harmonic and anharmonic approximations are shown in Fig.~S9(a)~\cite{refSM}. 
The negligible identical \(\lambda_v\) values further confirm that quantum fluctuations exert minimal influence on SrB\textsubscript{3}C\textsubscript{3}.

Quantum anharmonic effects have minimal impact on the anisotropy of \(\lambda_{nk}\) and \(\Delta_{nk}\), with \(\lambda_{nk}(\text{max}) - \lambda_{nk}(\text{min}) = 0.24\) (anharmonic) vs. 0.24 (harmonic) and \(\Delta_{nk}(\text{max}) - \Delta_{nk}(\text{min}) = 2.54\) meV (anharmonic) vs.~2.64 meV (harmonic) (see Figs.~\hyperref[fig:ph]{\ref*{fig:ph}(h)} and ~\hyperref[fig:ph]{\ref*{fig:ph}(i)}). 
The anisotropy of SrB\textsubscript{3}C\textsubscript{3} is comparable to that observed in hole-doped (BN)\textsubscript{6} and NH\textsubscript{4}B\textsubscript{2}C\textsubscript{8}~\cite{ref39,ref42}. 
Fig.~\hyperref[fig:ph]{\ref*{fig:ph}(j)} shows that anharmonicity has a negligible effect on the temperature dependence of the superconducting gap \(\Delta\) of SrB\textsubscript{3}C\textsubscript{3}, with $T_\text{c} = 44$ K (anharmonic) vs. 44.81 K (harmonic) at the Morel-Anderson pseudopotential \(\mu^* = 0.1\), in agreement with prior harmonic results~\cite{ref1,ref2}. 
The normalized superconducting density of states (SDOS) at 10 K for both approximations are in excellent agreement, exhibiting a typical single-gap feature in SrB\textsubscript{3}C\textsubscript{3} (see Fig.~\hyperref[fig:ph]{\ref*{fig:ph}(k)}). 
The single-gap feature is attributed to the $sp\textsuperscript{3}$-hybridized \(\sigma\) states, similar to those in other cage-like superconductors, including hole-doped C\textsubscript{6} and C\textsubscript{10} systems and NH\textsubscript{4}B\textsubscript{2}C\textsubscript{8}~\cite{ref38,ref42}.

\begin{figure}[htp]
	\centering
	\includegraphics[width=1\linewidth,angle=0]{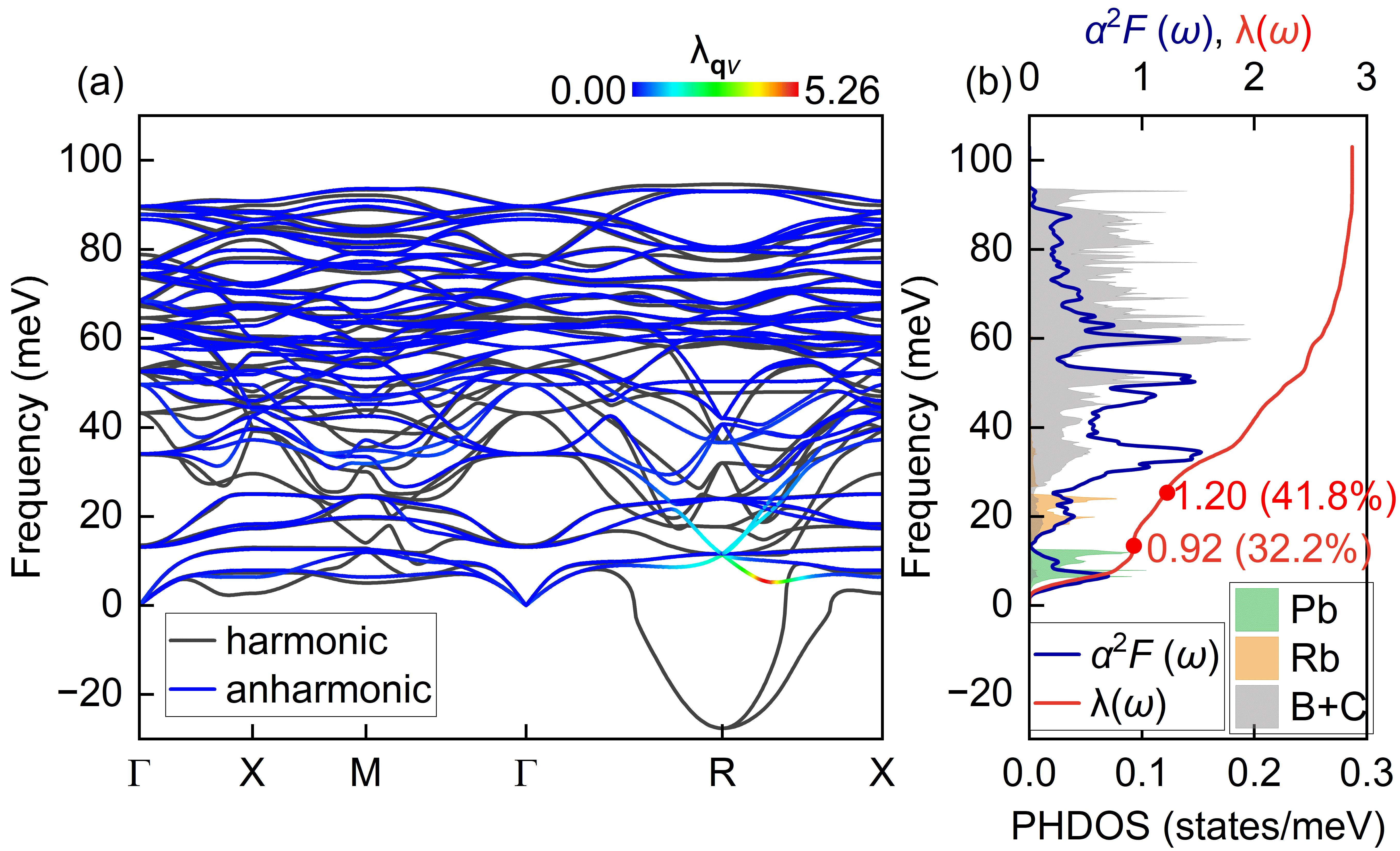}
	\caption{Anharmonic phonon properties and EPC of RbPbB$_6$C$_6$.
	(a) Phonon spectrum. The color map from blue to red represents the magnitude of $\lambda_{q\nu}$ for each phonon mode, with red denoting the highest coupling strengths.
    (b) Phonon density of states projected on Rb, Pb and the combined B+C atoms, Eliashberg spectral function $\alpha^2F(\omega)$, and integrated EPC strength $\lambda(\omega)$ in the anharmonic framework.
	}
    \label{fig:epc}
\end{figure}

As shown in Fig.~\hyperref[fig:epc]{\ref*{fig:epc}(a)}, the phonon dispersions for RbPbB\textsubscript{6}C\textsubscript{6} are presented, comparing the harmonic and anharmonic results. 
Quantum anharmonic effects eliminate the imaginary frequencies, confirming dynamical stability of RbPbB\textsubscript{6}C\textsubscript{6}. 
In a similar manner, PdH exhibits unstable phonon spectra in harmonic calculations but stabilizes when anharmonic effects are included~\cite{ref43}. 
Compared to SrB\textsubscript{3}C\textsubscript{3}, both the acoustic and low-lying optical branches (1\textendash 3) of RbPbB\textsubscript{6}C\textsubscript{6} are markedly softened. 
Similar to KSrB\textsubscript{6}C\textsubscript{6} and KPbB\textsubscript{6}C\textsubscript{6}~\cite{ref1}, which have an average metal oxidation state of +1.5, the softened modes near the R point show pronounced EPC contributions, particularly along the R\textendash X path. 
This softening is accompanied by a marked increase in the mode-resolved EPC strength \(\lambda_v\) (see Fig.~S9(b)~\cite{refSM}), resulting in a relatively high total EPC strength \(\lambda \approx 2.87\) for RbPbB\textsubscript{6}C\textsubscript{6}. 
Total EPC strength \(\lambda\) of RbPbB\textsubscript{6}C\textsubscript{6} exceeds that of SrB\textsubscript{3}C\textsubscript{3} (0.96), KPbB\textsubscript{6}C\textsubscript{6} (2.67)~\cite{ref1}, and Fmmm SrNH\textsubscript{4}B\textsubscript{6}C\textsubscript{6} (2.01)~\cite{ref44}. 
In Fig.~\hyperref[fig:epc]{\ref*{fig:epc}(b)}, the PHDOS for Pb, Rb, and B+C atoms, the Eliashberg function \(\alpha^2 F(\omega)\), and the integrated EPC strength \(\lambda(\omega)\) are shown as a function of frequency. 
The acoustic branches are primarily driven by Pb vibrations, while the low-lying optical branches (1\textendash 3) are dominated by Rb motions, consistent with their respective atomic masses. 
For \(\alpha^2 F(\omega)\) and \(\lambda(\omega)\), distinct contributions from different atoms (Pb, Rb, B + C) are observed. 
The first peak of \(\alpha^2 F(\omega)\) and the initial increase in \(\lambda(\omega)\) originate from coupling between Pb vibrations and electrons near the Fermi surface. 
Notably, the second phase of \(\lambda(\omega)\) is not associated with Rb motions. 
Although Rb dominates the PHDOS in the range 13.39\textendash 25.13 meV, the large \(\lambda_{qv}\) near the R point is primarily due to B and C vibrations, which are strongly coupled with the $sp\textsuperscript{3}$ hybridized \(\sigma\)-electrons (see Figs.~\hyperref[fig:epc]{\ref*{fig:epc}(a)} and S5(b)~\cite{refSM}). 
The higher optical branches (4\textendash 42) are mainly attributed to B and C vibrations, which contribute to the multiple peaks of \(\alpha^2 F(\omega)\) and the continuous increase in \(\lambda(\omega)\) above 25.13 meV.

\begin{figure}[htp]
	\centering
	\includegraphics[width=1\linewidth,angle=0]{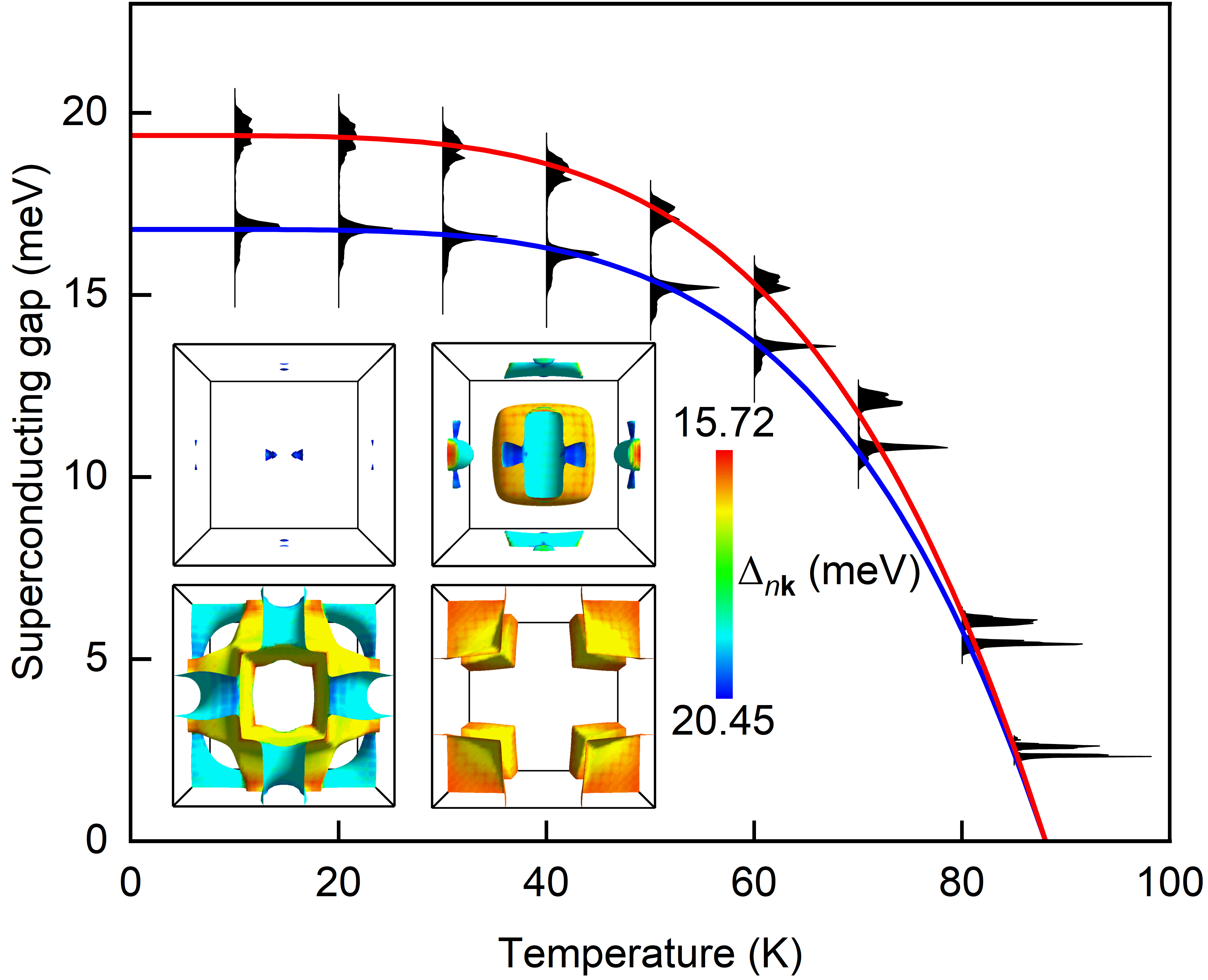}
	\caption{
	Temperature dependence of the superconducting gap $\Delta$ for RbPbB\textsubscript{6}C\textsubscript{6}, obtained by solving the anisotropic Migdal-Eliashberg equation at \(\mu^* = 0.1\).
	The inset shows the band- and momentum-resolved superconducting gap \(\Delta_{nk}\) on the Fermi surface at 10 K, with the color scale indicating \(\Delta_{nk}\) values ranging from 15.72 meV (blue) to 20.45 meV (red).
	}
	\label{fig:gap}
\end{figure}

Figure~\ref{fig:gap} presents the temperature dependence of the superconducting gap \(\Delta\) for RbPbB\textsubscript{6}C\textsubscript{6} at \(\mu^* = 0.1\). 
The two superconducting gaps decrease with increasing temperature, closing simultaneously at $T_\text{c} = 88$ K. 
Anharmonic $T_\text{c}$ of RbPbB\textsubscript{6}C\textsubscript{6} exceeds that of SrB\textsubscript{3}C\textsubscript{3} (40\textendash 45 K)~\cite{ref2,ref15,ref40,ref41}, RbYbB\textsubscript{6}C\textsubscript{6} (67 K)~\cite{ref45}, and Fmmm SrNH\textsubscript{4}B\textsubscript{6}C\textsubscript{6} (85 K)~\cite{ref44}, comparable to that of hydrogen-modified alloy (42\textendash 77 K)~\cite{ref46,ref47,ref48}. 
At 10 K, the average values of the two superconducting gaps are 16.61 meV and 14.01 meV, corresponding to two distinct peaks in the SDOS, as shown in Fig.~S10~\cite{refSM}. 
The inset of Fig.~\ref{fig:gap} also presents \(\Delta_{nk}\) on the Fermi surface (FS) at 10 K, with \(\Delta_{nk}\) varying between 15.72 meV and 20.45 meV. 
This observation confirms that RbPbB\textsubscript{6}C\textsubscript{6} is a two-gap superconductor, similar to MgB\textsubscript{2} and Janus MoSH~\cite{ref49,ref50}. 
For RbPbB$_6$C$_6$, the higher C/B ratio in the density of states at the Fermi level compared to SrB$_3$C$_3$, weakens $sp^3$ hybridization, leading to electronic states dominated by C and B characteristics, each opening a distinct gap.
Importantly, larger values of \(\Delta_{nk}\) on the FS are found to correlate with higher EPC strength \(\lambda_{nk}\), as shown in Fig.~S11~\cite{refSM}, further supporting the phonon-mediated nature of superconductivity in RbPbB\textsubscript{6}C\textsubscript{6}. 
The largest gap, observed on FS4, is attributed to Pb phonon modes with strong EPC along the R\textendash X direction.

\section{CONCLUSIONS}
In conclusion, we demonstrate the crucial role of quantum effects in determining the stability and superconductivity of binary-guest carbon-boron clathrates XYB\textsubscript{6}C\textsubscript{6}, challenging harmonic-level predictions and evidencing flaws in previous studies. 
Quantum effects are induced by the guest atoms, with their magnitude depending on the atomic species. 
For SrB\textsubscript{3}C\textsubscript{3}, quantum effects have little impact on its dynamical stability and $T_\text{c}$. 
In contrast, pronounced quantum effects arise when Pb is one of the X/Y elements. 
Including quantum effects stabilizes RbPbB\textsubscript{6}C\textsubscript{6} and raises $T_\text{c}$ to 88 K, nearly twice that of SrB\textsubscript{3}C\textsubscript{3} (44 K). 
The enhanced $T_\text{c}$ originates from two key factors: (1) the lower oxidation state of RbPb, which shifts the Fermi level downward, thereby increasing DOS at the Fermi level, and (2) the softened phonon modes, both of which collectively strengthens the electron-phonon coupling of 2.87. 
These findings provide a reliable framework for assessing dynamical stability and superconducting $T_\text{c}$ in carbon-boron clathrates exhibiting strong quantum anharmonic effects.

\section{ACKNOWLEDGMENTS}
We acknowledge the support by National Natural Science Foundation of China (No.52227813, No.11204053 and No.11074059).

\bibliography{references.bib}


\end{document}